
%
%
%
%
%
%
%
%
%
\def\standardrisposta{s }\def\reducedrisposta{r }
\def\doublerisposta{d }\def\cartarisposta{e }\def\amsrisposta{y }
\newcount\ingrandimento \newcount\sinnota \newcount\dimnota
\newcount\unoduecol \newdimen\collhsize \newdimen\tothsize
\newdimen\fullhsize \newcount\controllorisposta \sinnota=1
\newskip\infralinea  \global\controllorisposta=0
\message{ ********    Welcome to PANDA macros (Plain TeX, AP, 1991)}
\message{ ******** }
\message{       You'll have to answer a few questions in lowercase.}
\message{>  Do you want it in double-page (d), reduced (r)}
\message{or standard format (s) ? }\read-1 to\risposta
\message{>  Do you want it in USA A4 (u) or EUROPEAN A4 (e)}
\message{paper size ? }\read-1 to\srisposta
\message{>  Do you have AMSFonts 2.0 (math) fonts (y/n) ? }
\read-1 to\arisposta
%
%
%
%
%
\ifx\risposta\standardrisposta \ingrandimento=1200
\message{>> This will come out UNREDUCED << }
\dimnota=2 \unoduecol=1 \global\controllorisposta=1 \fi
\ifx\risposta\reducedrisposta \ingrandimento=1095 \dimnota=1
\unoduecol=1  \global\controllorisposta=1
\message{>> This will come out REDUCED << } \fi
\ifx\risposta\doublerisposta \ingrandimento=1000 \dimnota=2
\unoduecol=2  \global\controllorisposta=1
\message{>> You must print this in LANDSCAPE orientation << } \fi
\ifnum\controllorisposta=0  \ingrandimento=1200
\message{>>> ERROR IN INPUT, I ASSUME STANDARD UNREDUCED FORMAT <<< }
\dimnota=2 \unoduecol=1 \fi
\magnification=\ingrandimento
%
%
%
%
\newdimen\eucolumnsize \newdimen\eudoublehsize \newdimen\eudoublevsize
\newdimen\uscolumnsize \newdimen\usdoublehsize \newdimen\usdoublevsize
\newdimen\eusinglehsize \newdimen\eusinglevsize \newdimen\ussinglehsize
\newskip\standardbaselineskip \newdimen\ussinglevsize
\newskip\reducedbaselineskip \newskip\doublebaselineskip
\eucolumnsize=12.0truecm    
\eudoublehsize=25.5truecm   
\eudoublevsize=6.5truein    
\uscolumnsize=4.4truein     
\usdoublehsize=9.4truein    
\usdoublevsize=6.8truein    
\eusinglehsize=6.5truein    
\eusinglevsize=24truecm     
\ussinglehsize=6.5truein    
\ussinglevsize=8.9truein    
\standardbaselineskip=16pt  
\reducedbaselineskip=14pt   
\doublebaselineskip=12pt    
%
%
\def\Portoffset{}
\def\Landoffset{}
%
%

%
\tolerance=10000
\parskip 0pt plus 2pt  
%
%
\ifx\risposta\standardrisposta \infralinea=\standardbaselineskip \fi
\ifx\risposta\reducedrisposta  \infralinea=\reducedbaselineskip \fi
\ifx\risposta\doublerisposta   \infralinea=\doublebaselineskip \fi
\ifnum\controllorisposta=0    \infralinea=\standardbaselineskip \fi
\ifx\risposta\doublerisposta   \Landoffset \else \Portoffset \fi
\ifx\risposta\doublerisposta \ifx\srisposta\cartarisposta
\tothsize=\eudoublehsize \collhsize=\eucolumnsize
\vsize=\eudoublevsize  \else  \tothsize=\usdoublehsize
\collhsize=\uscolumnsize \vsize=\usdoublevsize \fi \else
\ifx\srisposta\cartarisposta \tothsize=\eusinglehsize
\vsize=\eusinglevsize \else  \tothsize=\ussinglehsize
\vsize=\ussinglevsize \fi \collhsize=4.4truein \fi
%
%
%
%
\newcount\contaeuler \newcount\contacyrill \newcount\contaams
\font\ninerm=cmr9  \font\eightrm=cmr8  \font\sixrm=cmr6
\font\ninei=cmmi9  \font\eighti=cmmi8  \font\sixi=cmmi6
\font\ninesy=cmsy9  \font\eightsy=cmsy8  \font\sixsy=cmsy6
\font\ninebf=cmbx9  \font\eightbf=cmbx8  \font\sixbf=cmbx6
\font\ninett=cmtt9  \font\eighttt=cmtt8  \font\nineit=cmti9
\font\eightit=cmti8 \font\ninesl=cmsl9  \font\eightsl=cmsl8
\skewchar\ninei='177 \skewchar\eighti='177 \skewchar\sixi='177
\skewchar\ninesy='60 \skewchar\eightsy='60 \skewchar\sixsy='60
\hyphenchar\ninett=-1 \hyphenchar\eighttt=-1 \hyphenchar\tentt=-1
\def\bfmath{\cmmib}                 
\font\tencmmib=cmmib10  \newfam\cmmibfam  \skewchar\tencmmib='177
\font\tencmbsy=cmbsy10  \newfam\cmbsyfam  \skewchar\tencmbsy='60
\def\scaps{\cmcsc}                 
\font\tencmcsc=cmcsc10  \newfam\cmcscfam
\ifnum\ingrandimento=1095

\font\capsone=cmcsc10 at 10.95pt 

\else

\font\capsone=cmcsc10 at 12pt 
\fi

\def\ttaarr{\bf}                
\def\ppaarr{\sl}                

%
%
%
\newfam\eufmfam \newfam\msamfam \newfam\msbmfam \newfam\eufbfam
\def\Loadeulerfonts{\global\contaeuler=1 \ifx\arisposta\amsrisposta
\font\teneufm=eufm10              
\font\eighteufm=eufm8 \font\nineeufm=eufm9 \font\sixeufm=eufm6
\font\seveneufm=eufm7  \font\fiveeufm=eufm5
\font\teneufb=eufb10              
\font\eighteufb=eufb8 \font\nineeufb=eufb9 \font\sixeufb=eufb6
\font\seveneufb=eufb7  \font\fiveeufb=eufb5
\font\teneurm=eurm10              
\font\eighteurm=eurm8 \font\nineeurm=eurm9
\font\teneurb=eurb10              
\font\eighteurb=eurb8 \font\nineeurb=eurb9
\font\teneusm=eusm10              
\font\eighteusm=eusm8 \font\nineeusm=eusm9
\font\teneusb=eusb10              
\font\eighteusb=eusb8 \font\nineeusb=eusb9
\else \def\eufm{\tt} \def\eufb{\tt} \def\eurm{\tt} \def\eurb{\tt}
\def\eusm{\tt} \def\eusb{\tt}    \fi}
\def\loadeuler{\Loadeulerfonts\tenpoint}
\def\loadamsmath{\global\contaams=1 \ifx\arisposta\amsrisposta
\font\tenmsam=msam10 \font\ninemsam=msam9 \font\eightmsam=msam8
\font\sevenmsam=msam7 \font\sixmsam=msam6 \font\fivemsam=msam5
\font\tenmsbm=msbm10 \font\ninemsbm=msbm9 \font\eightmsbm=msbm8
\font\sevenmsbm=msbm7 \font\sixmsbm=msbm6 \font\fivemsbm=msbm5
\else \def\msbm{\bf} \fi \def\Bbb{\msbm} \def\symbl{\msam} \tenpoint}
\def\loadcyrill{\global\contacyrill=1 \ifx\arisposta\amsrisposta
\font\tenwncyr=wncyr10 \font\ninewncyr=wncyr9 \font\eightwncyr=wncyr8
\font\tenwncyb=wncyr10 \font\ninewncyb=wncyr9 \font\eightwncyb=wncyr8
\font\tenwncyi=wncyr10 \font\ninewncyi=wncyr9 \font\eightwncyi=wncyr8
\else \def\cyrill{\sl} \def\cyrilb{\sl} \def\cyrili{\sl} \fi\tenpoint}
\ifx\arisposta\amsrisposta
\font\sevenex=cmex7               
\font\eightex=cmex8  \font\nineex=cmex9
\font\ninecmmib=cmmib9   \font\eightcmmib=cmmib8
\font\sevencmmib=cmmib7 \font\sixcmmib=cmmib6
\font\fivecmmib=cmmib5   \skewchar\ninecmmib='177
\skewchar\eightcmmib='177  \skewchar\sevencmmib='177
\skewchar\sixcmmib='177   \skewchar\fivecmmib='177
\font\ninecmbsy=cmbsy9    \font\eightcmbsy=cmbsy8
\font\sevencmbsy=cmbsy7  \font\sixcmbsy=cmbsy6
\font\fivecmbsy=cmbsy5   \skewchar\ninecmbsy='60
\skewchar\eightcmbsy='60  \skewchar\sevencmbsy='60
\skewchar\sixcmbsy='60    \skewchar\fivecmbsy='60
\font\ninecmcsc=cmcsc9    \font\eightcmcsc=cmcsc8     \else
\def\cmmib{\fam\cmmibfam\tencmmib}\textfont\cmmibfam=\tencmmib
\scriptfont\cmmibfam=\tencmmib \scriptscriptfont\cmmibfam=\tencmmib
\def\cmbsy{\fam\cmbsyfam\tencmbsy} \textfont\cmbsyfam=\tencmbsy
\scriptfont\cmbsyfam=\tencmbsy \scriptscriptfont\cmbsyfam=\tencmbsy
\scriptfont\cmcscfam=\tencmcsc \scriptscriptfont\cmcscfam=\tencmcsc
\def\cmcsc{\fam\cmcscfam\tencmcsc} \textfont\cmcscfam=\tencmcsc \fi
\catcode`@=11
\newskip\ttglue
\gdef\tenpoint{\def\rm{\fam0\tenrm}
  \textfont0=\tenrm \scriptfont0=\sevenrm \scriptscriptfont0=\fiverm
  \textfont1=\teni \scriptfont1=\seveni \scriptscriptfont1=\fivei
  \textfont2=\tensy \scriptfont2=\sevensy \scriptscriptfont2=\fivesy
  \textfont3=\tenex \scriptfont3=\tenex \scriptscriptfont3=\tenex
  \def\mcal{\fam2 \tensy}  \def\mmit{\fam1 \teni}
  \textfont\itfam=\tenit \def\it{\fam\itfam\tenit}
  \textfont\slfam=\tensl \def\sl{\fam\slfam\tensl}
  \textfont\ttfam=\tentt \scriptfont\ttfam=\eighttt
  \scriptscriptfont\ttfam=\eighttt  \def\tt{\fam\ttfam\tentt}
  \textfont\bffam=\tenbf \scriptfont\bffam=\sevenbf
  \scriptscriptfont\bffam=\fivebf \def\bf{\fam\bffam\tenbf}
     \ifx\arisposta\amsrisposta    \ifnum\contaeuler=1
  \textfont\eufmfam=\teneufm \scriptfont\eufmfam=\seveneufm
  \scriptscriptfont\eufmfam=\fiveeufm \def\eufm{\fam\eufmfam\teneufm}
  \textfont\eufbfam=\teneufb \scriptfont\eufbfam=\seveneufb
  \scriptscriptfont\eufbfam=\fiveeufb \def\eufb{\fam\eufbfam\teneufb}
  \def\eurm{\teneurm} \def\eurb{\teneurb} \def\eusm{\teneusm}
  \def\eusb{\teneusb}    \fi    \ifnum\contaams=1
  \textfont\msamfam=\tenmsam \scriptfont\msamfam=\sevenmsam
  \scriptscriptfont\msamfam=\fivemsam \def\msam{\fam\msamfam\tenmsam}
  \textfont\msbmfam=\tenmsbm \scriptfont\msbmfam=\sevenmsbm
  \scriptscriptfont\msbmfam=\fivemsbm \def\msbm{\fam\msbmfam\tenmsbm}
     \fi      \ifnum\contacyrill=1     \def\cyrill{\tenwncyr}
  \def\cyrilb{\tenwncyb}  \def\cyrili{\tenwncyi}         \fi
  \textfont3=\tenex \scriptfont3=\sevenex \scriptscriptfont3=\sevenex
  \def\cmmib{\fam\cmmibfam\tencmmib} \scriptfont\cmmibfam=\sevencmmib
  \textfont\cmmibfam=\tencmmib  \scriptscriptfont\cmmibfam=\fivecmmib
  \def\cmbsy{\fam\cmbsyfam\tencmbsy} \scriptfont\cmbsyfam=\sevencmbsy
  \textfont\cmbsyfam=\tencmbsy  \scriptscriptfont\cmbsyfam=\fivecmbsy
  \def\cmcsc{\fam\cmcscfam\tencmcsc} \scriptfont\cmcscfam=\eightcmcsc
  \textfont\cmcscfam=\tencmcsc \scriptscriptfont\cmcscfam=\eightcmcsc
     \fi            \tt \ttglue=.5em plus.25em minus.15em
  \normalbaselineskip=12pt
  \setbox\strutbox=\hbox{\vrule height8.5pt depth3.5pt width0pt}
  \let\sc=\eightrm \let\big=\tenbig   \normalbaselines
  \baselineskip=\infralinea  \rm}
\gdef\ninepoint{\def\rm{\fam0\ninerm}
  \textfont0=\ninerm \scriptfont0=\sixrm \scriptscriptfont0=\fiverm
  \textfont1=\ninei \scriptfont1=\sixi \scriptscriptfont1=\fivei
  \textfont2=\ninesy \scriptfont2=\sixsy \scriptscriptfont2=\fivesy
  \textfont3=\tenex \scriptfont3=\tenex \scriptscriptfont3=\tenex
  \def\mcal{\fam2 \ninesy}  \def\mmit{\fam1 \ninei}
  \textfont\itfam=\nineit \def\it{\fam\itfam\nineit}
  \textfont\slfam=\ninesl \def\sl{\fam\slfam\ninesl}
  \textfont\ttfam=\ninett \scriptfont\ttfam=\eighttt
  \scriptscriptfont\ttfam=\eighttt \def\tt{\fam\ttfam\ninett}
  \textfont\bffam=\ninebf \scriptfont\bffam=\sixbf
  \scriptscriptfont\bffam=\fivebf \def\bf{\fam\bffam\ninebf}
     \ifx\arisposta\amsrisposta  \ifnum\contaeuler=1
  \textfont\eufmfam=\nineeufm \scriptfont\eufmfam=\sixeufm
  \scriptscriptfont\eufmfam=\fiveeufm \def\eufm{\fam\eufmfam\nineeufm}
  \textfont\eufbfam=\nineeufb \scriptfont\eufbfam=\sixeufb
  \scriptscriptfont\eufbfam=\fiveeufb \def\eufb{\fam\eufbfam\nineeufb}
  \def\eurm{\nineeurm} \def\eurb{\nineeurb} \def\eusm{\nineeusm}
  \def\eusb{\nineeusb}     \fi   \ifnum\contaams=1
  \textfont\msamfam=\ninemsam \scriptfont\msamfam=\sixmsam
  \scriptscriptfont\msamfam=\fivemsam \def\msam{\fam\msamfam\ninemsam}
  \textfont\msbmfam=\ninemsbm \scriptfont\msbmfam=\sixmsbm
  \scriptscriptfont\msbmfam=\fivemsbm \def\msbm{\fam\msbmfam\ninemsbm}
     \fi       \ifnum\contacyrill=1     \def\cyrill{\ninewncyr}
  \def\cyrilb{\ninewncyb}  \def\cyrili{\ninewncyi}         \fi
  \textfont3=\nineex \scriptfont3=\sevenex \scriptscriptfont3=\sevenex
  \def\cmmib{\fam\cmmibfam\ninecmmib}  \textfont\cmmibfam=\ninecmmib
  \scriptfont\cmmibfam=\sixcmmib \scriptscriptfont\cmmibfam=\fivecmmib
  \def\cmbsy{\fam\cmbsyfam\ninecmbsy}  \textfont\cmbsyfam=\ninecmbsy
  \scriptfont\cmbsyfam=\sixcmbsy \scriptscriptfont\cmbsyfam=\fivecmbsy
  \def\cmcsc{\fam\cmcscfam\ninecmcsc} \scriptfont\cmcscfam=\eightcmcsc
  \textfont\cmcscfam=\ninecmcsc \scriptscriptfont\cmcscfam=\eightcmcsc
     \fi            \tt \ttglue=.5em plus.25em minus.15em
  \normalbaselineskip=11pt
  \setbox\strutbox=\hbox{\vrule height8pt depth3pt width0pt}
  \let\sc=\sevenrm \let\big=\ninebig \normalbaselines\rm}
\gdef\eightpoint{\def\rm{\fam0\eightrm}
  \textfont0=\eightrm \scriptfont0=\sixrm \scriptscriptfont0=\fiverm
  \textfont1=\eighti \scriptfont1=\sixi \scriptscriptfont1=\fivei
  \textfont2=\eightsy \scriptfont2=\sixsy \scriptscriptfont2=\fivesy
  \textfont3=\tenex \scriptfont3=\tenex \scriptscriptfont3=\tenex
  \def\mcal{\fam2 \eightsy}  \def\mmit{\fam1 \eighti}
  \textfont\itfam=\eightit \def\it{\fam\itfam\eightit}
  \textfont\slfam=\eightsl \def\sl{\fam\slfam\eightsl}
  \textfont\ttfam=\eighttt \scriptfont\ttfam=\eighttt
  \scriptscriptfont\ttfam=\eighttt \def\tt{\fam\ttfam\eighttt}
  \textfont\bffam=\eightbf \scriptfont\bffam=\sixbf
  \scriptscriptfont\bffam=\fivebf \def\bf{\fam\bffam\eightbf}
     \ifx\arisposta\amsrisposta   \ifnum\contaeuler=1
  \textfont\eufmfam=\eighteufm \scriptfont\eufmfam=\sixeufm
  \scriptscriptfont\eufmfam=\fiveeufm \def\eufm{\fam\eufmfam\eighteufm}
  \textfont\eufbfam=\eighteufb \scriptfont\eufbfam=\sixeufb
  \scriptscriptfont\eufbfam=\fiveeufb \def\eufb{\fam\eufbfam\eighteufb}
  \def\eurm{\eighteurm} \def\eurb{\eighteurb} \def\eusm{\eighteusm}
  \def\eusb{\eighteusb}       \fi    \ifnum\contaams=1
  \textfont\msamfam=\eightmsam \scriptfont\msamfam=\sixmsam
  \scriptscriptfont\msamfam=\fivemsam \def\msam{\fam\msamfam\eightmsam}
  \textfont\msbmfam=\eightmsbm \scriptfont\msbmfam=\sixmsbm
  \scriptscriptfont\msbmfam=\fivemsbm \def\msbm{\fam\msbmfam\eightmsbm}
     \fi       \ifnum\contacyrill=1     \def\cyrill{\eightwncyr}
  \def\cyrilb{\eightwncyb}  \def\cyrili{\eightwncyi}         \fi
  \textfont3=\eightex \scriptfont3=\sevenex \scriptscriptfont3=\sevenex
  \def\cmmib{\fam\cmmibfam\eightcmmib}  \textfont\cmmibfam=\eightcmmib
  \scriptfont\cmmibfam=\sixcmmib \scriptscriptfont\cmmibfam=\fivecmmib
  \def\cmbsy{\fam\cmbsyfam\eightcmbsy}  \textfont\cmbsyfam=\eightcmbsy
  \scriptfont\cmbsyfam=\sixcmbsy \scriptscriptfont\cmbsyfam=\fivecmbsy
  \def\cmcsc{\fam\cmcscfam\eightcmcsc} \scriptfont\cmcscfam=\eightcmcsc
  \textfont\cmcscfam=\eightcmcsc \scriptscriptfont\cmcscfam=\eightcmcsc
     \fi             \tt \ttglue=.5em plus.25em minus.15em
  \normalbaselineskip=9pt
  \setbox\strutbox=\hbox{\vrule height7pt depth2pt width0pt}
  \let\sc=\sixrm \let\big=\eightbig \normalbaselines\rm}
\gdef\tenbig#1{{\hbox{$\left#1\vbox to8.5pt{}\right.\n@space$}}}
\gdef\ninebig#1{{\hbox{$\textfont0=\tenrm\textfont2=\tensy
   \left#1\vbox to7.25pt{}\right.\n@space$}}}
\gdef\eightbig#1{{\hbox{$\textfont0=\ninerm\textfont2=\ninesy
   \left#1\vbox to6.5pt{}\right.\n@space$}}}
\def\alternativefont#1#2{\ifx\arisposta\amsrisposta \relax \else
\xdef#1{#2} \fi}
\global\contaeuler=0 \global\contacyrill=0 \global\contaams=0
%
%
%
%
\newbox\fotlinebb \newbox\hedlinebb \newbox\leftcolumn
\gdef\makeheadline{\vbox to 0pt{\vskip-22.5pt
     \fullline{\vbox to8.5pt{}\the\headline}\vss}\nointerlineskip}
\gdef\makehedlinebb{\vbox to 0pt{\vskip-22.5pt
     \fullline{\vbox to8.5pt{}\copy\hedlinebb\hfil
     \line{\hfill\the\headline\hfill}}\vss} \nointerlineskip}
\gdef\makefootline{\baselineskip=24pt \fullline{\the\footline}}
\gdef\makefotlinebb{\baselineskip=24pt
    \fullline{\copy\fotlinebb\hfil\line{\hfill\the\footline\hfill}}}
\gdef\doubleformat{\shipout\vbox{\makehedlinebb
     \fullline{\box\leftcolumn\hfil\columnbox}\makefotlinebb}
     \advancepageno}

\gdef\columnbox{\leftline{\pagebody}}
\gdef\line#1{\hbox to\hsize{\hskip\leftskip#1\hskip\rightskip}}
\gdef\fullline#1{\hbox to\fullhsize{\hskip\leftskip{#1}%
\hskip\rightskip}}
\gdef\footnote#1{\let\@sf=\empty
         \ifhmode\edef\#sf{\spacefactor=\the\spacefactor}\/\fi
         #1\@sf\vfootnote{#1}}
\gdef\vfootnote#1{\insert\footins\bgroup
         \ifnum\dimnota=1  \eightpoint\fi
         \ifnum\dimnota=2  \ninepoint\fi
         \ifnum\dimnota=0  \tenpoint\fi
         \interlinepenalty=\interfootnotelinepenalty
         \splittopskip=\ht\strutbox
         \splitmaxdepth=\dp\strutbox \floatingpenalty=20000
         \leftskip=\oldssposta \rightskip=\olddsposta
         \spaceskip=0pt \xspaceskip=0pt
         \ifnum\sinnota=0   \textindent{#1}\fi
         \ifnum\sinnota=1   \item{#1}\fi
         \footstrut\futurelet\next\fo@t}
\gdef\fo@t{\ifcat\bgroup\noexpand\next \let\next\f@@t
             \else\let\next\f@t\fi \next}
\gdef\f@@t{\bgroup\aftergroup\@foot\let\next}
\gdef\f@t#1{#1\@foot} \gdef\@foot{\strut\egroup}
\gdef\footstrut{\vbox to\splittopskip{}}
\skip\footins=\bigskipamount
\count\footins=1000  \dimen\footins=8in
\catcode`@=12
\tenpoint      
\newskip\olddsposta \newskip\oldssposta
\global\oldssposta=\leftskip \global\olddsposta=\rightskip

\gdef\yespagenumbers{\footline={\hss\tenrm\folio\hss}}
\gdef\ciao{\par\vfill\supereject \ifnum\unoduecol=2
     \if R\lrcol  \headline={}\nopagenumbers\null\vfill\eject
     \fi\fi \end}

\ifnum\unoduecol=1 \hsize=\tothsize   \fullhsize=\tothsize \fi
\ifnum\unoduecol=2 \hsize=\collhsize  \fullhsize=\tothsize \fi
\global\let\lrcol=L
\ifnum\unoduecol=1 \output{\plainoutput}\fi
\ifnum\unoduecol=2 \output{\if L\lrcol
       \global\setbox\leftcolumn=\columnbox
       \global\setbox\fotlinebb=\line{\hfill\the\footline\hfill}
       \global\setbox\hedlinebb=\line{\hfill\the\headline\hfill}
       \advancepageno
      \global\let\lrcol=R \else \doubleformat \global\let\lrcol=L \fi
       \ifnum\outputpenalty>-20000 \else\dosupereject\fi}\fi
\def\ifdoublepage{\ifnum\unoduecol=2 }
\def\filldots{\leaders\hbox to 1em{\hss.\hss}\hfill}
\def\inquadrb#1 {\vbox {\hrule  \hbox{\vrule \vbox {\vskip .2cm
    \hbox {\ #1\ } \vskip .2cm } \vrule  }  \hrule} }

\def\newline{\hfil\break}
\def\jump{\vskip\baselineskip} \newskip\iinnffrr
\def\sjump{\iinnffrr=\baselineskip
          \divide\iinnffrr by 2 \vskip\iinnffrr}
\def\bjump{\vskip\baselineskip \vskip\baselineskip}
\newcount\nmbnota  \def\clearnmbnota{\global\nmbnota=0}
\def\note#1{\global\advance\nmbnota by 1
    \footnote{$^{\the\nmbnota}$}{#1}}  \clearnmbnota
\def\setnote#1{\expandafter\xdef\csname#1\endcsname{\the\nmbnota}}
\newcount\nbmfig  \def\clearnbmfig{\global\nbmfig=0}
\gdef\figure{\global\advance\nbmfig by 1
      {\rm fig. \the\nbmfig}}   \clearnbmfig
\def\setfig#1{\expandafter\xdef\csname#1\endcsname{fig. \the\nbmfig}}
 \def\endformula{\eqno\numero $$}
 \def\efr{\endformula}
\newcount\frmcount \def\clearfrmcount{\global\frmcount=0}
\def\numero{\global\advance\frmcount by 1   \ifnum\indappcount=0
  {\ifnum\cpcount <1 {\hbox{\rm (\the\frmcount )}}  \else
  {\hbox{\rm (\the\cpcount .\the\frmcount )}} \fi}  \else
  {\hbox{\rm (\applett .\the\frmcount )}} \fi}
\def\nameformula#1{\global\advance\frmcount by 1%
\ifnum\draftnum=0  {\ifnum\indappcount=0%
{\ifnum\cpcount<1\xdef\spzzttrra{(\the\frmcount )}%
\else\xdef\spzzttrra{(\the\cpcount .\the\frmcount )}\fi}%
\else\xdef\spzzttrra{(\applett .\the\frmcount )}\fi}%
\else\xdef\spzzttrra{(#1)}\fi%
\expandafter\xdef\csname#1\endcsname{\spzzttrra}
\eqno \ifnum\draftnum=0 {\ifnum\indappcount=0
  {\ifnum\cpcount <1 {\hbox{\rm (\the\frmcount )}}  \else
  {\hbox{\rm (\the\cpcount .\the\frmcount )}}\fi}   \else
  {\hbox{\rm (\applett .\the\frmcount )}} \fi} \else (#1) \fi $$}
\def\nfr{\nameformula}    
\def\nameali#1{\global\advance\frmcount by 1%
\ifnum\draftnum=0  {\ifnum\indappcount=0%
{\ifnum\cpcount<1\xdef\spzzttrra{(\the\frmcount )}%
\else\xdef\spzzttrra{(\the\cpcount .\the\frmcount )}\fi}%
\else\xdef\spzzttrra{(\applett .\the\frmcount )}\fi}%
\else\xdef\spzzttrra{(#1)}\fi%
\expandafter\xdef\csname#1\endcsname{\spzzttrra}
  \ifnum\draftnum=0  {\ifnum\indappcount=0
  {\ifnum\cpcount <1 {\hbox{\rm (\the\frmcount )}}  \else
  {\hbox{\rm (\the\cpcount .\the\frmcount )}}\fi}   \else
  {\hbox{\rm (\applett .\the\frmcount )}} \fi} \else (#1) \fi}
\clearfrmcount
\newcount\cpcount \def\clearcpcount{\global\cpcount=0}
\newcount\subcpcount \def\clearsubcpcount{\global\subcpcount=0}
\newcount\appcount \def\clearappcount{\global\appcount=0}
\newcount\indappcount \def\clearindappcount{\indappcount=0}
\newcount\sottoparcount 

\def\applett{\ifcase\appcount  \or {A}\or {B}\or {C}\or
{D}\or {E}\or {F}\or {G}\or {H}\or {I}\or {J}\or {K}\or {L}\or
{M}\or {N}\or {O}\or {P}\or {Q}\or {R}\or {S}\or {T}\or {U}\or
{V}\or {W}\or {X}\or {Y}\or {Z}\fi
             \ifnum\appcount<0
    \message{>>  ERROR: counter \appcount out of range <<}\fi
             \ifnum\appcount>26
   \message{>>  ERROR: counter \appcount out of range <<}\fi}
\clearappcount  \clearindappcount
\newcount\connttrre  \def\clearconnttrre{\global\connttrre=0}
\newcount\countref  \def\clearcountref{\global\countref=0}
\clearcountref
\def\chapter#1{\global\advance\cpcount by 1 \clearfrmcount
                 \goodbreak\null\vbox{\jump\nobreak
                 \clearsubcpcount\clearindappcount
                 \itemitem{\ttaarr\the\cpcount .\qquad}{\ttaarr #1}
                 \par\nobreak\jump\sjump}\nobreak}
\def\section#1{\global\advance\subcpcount by 1 \goodbreak\null
               \vbox{\sjump\nobreak\ifnum\indappcount=0
                 {\ifnum\cpcount=0 {\itemitem{\ppaarr
               .\the\subcpcount\quad\enskip\ }{\ppaarr #1}\par} \else
                 {\itemitem{\ppaarr\the\cpcount .\the\subcpcount\quad
                  \enskip\ }{\ppaarr #1} \par}  \fi}
                \else{\itemitem{\ppaarr\applett .\the\subcpcount\quad
                 \enskip\ }{\ppaarr #1}\par}\fi\nobreak\jump}\nobreak}
\clearsubcpcount
\def\appendix#1{\global\advance\appcount by 1 \clearfrmcount
                  \goodbreak\null\vbox{\jump\nobreak
                  \global\advance\indappcount by 1 \clearsubcpcount
                  \noindent{\ttaarr Appendix\ \applett\ }{\ttaarr #1}
                  \nobreak\jump\sjump}\nobreak}
\clearappcount \clearindappcount
\def\references{\goodbreak\null\vbox{\jump\nobreak
   \noindent{\ttaarr References} \nobreak\jump\sjump}\nobreak}

\clearcpcount\clearcountref
\def\acknowledgements{\goodbreak\null\vbox{\jump\nobreak
\centerline{{\ttaarr Acknowledgements}} \nobreak\jump\sjump}\nobreak}
\def\setchap#1{\ifnum\indappcount=0{\ifnum\subcpcount=0%
\xdef\spzzttrra{\the\cpcount}%
\else\xdef\spzzttrra{\the\cpcount .\the\subcpcount}\fi}
\else{\ifnum\subcpcount=0 \xdef\spzzttrra{\applett}%
\else\xdef\spzzttrra{\applett .\the\subcpcount}\fi}\fi
\expandafter\xdef\csname#1\endcsname{\spzzttrra}}
\newcount\draftnum \newcount\ppora   \newcount\ppminuti
\global\ppora=\time   \global\ppminuti=\time
\global\divide\ppora by 60  \draftnum=\ppora
\multiply\draftnum by 60    \global\advance\ppminuti by -\draftnum
\global\draftnum=0
\def\droggi{\number\day /\number\month /\number\year\ \the\ppora
:\the\ppminuti}

\global\draftnum=0
\def\draftcomment#1{\ifnum\draftnum=0 \relax \else {\ {\bf ***}\ #1\
{\bf ***}\ }\fi}

%
%
\catcode`@=11
\gdef\Ref#1{\expandafter\ifx\csname @rrxx@#1\endcsname\relax%
{\global\advance\countref by 1%
\ifnum\countref>200%
\message{>>> ERROR: maximum number of references exceeded <<<}%
\expandafter\xdef\csname @rrxx@#1\endcsname{0}\else%
\expandafter\xdef\csname @rrxx@#1\endcsname{\the\countref}\fi}\fi%
\ifnum\draftnum=0 \csname @rrxx@#1\endcsname \else#1\fi}
\gdef\beginref{\ifnum\draftnum=0  \gdef\Rref{\fairef}
\gdef\endref{\scriviref} \else\relax\fi}
\def\Reflab#1{[#1]}
\gdef\Rref#1#2{\item{\Reflab{#1}}{#2}}  \gdef\endref{\relax}
\newcount\conttemp
\gdef\fairef#1#2{\expandafter\ifx\csname @rrxx@#1\endcsname\relax
{\global\conttemp=0
\message{>>> ERROR: reference [#1] not defined <<<} } \else
{\global\conttemp=\csname @rrxx@#1\endcsname } \fi
\global\advance\conttemp by 50
\global\setbox\conttemp=\hbox{#2} }
\gdef\scriviref{\clearconnttrre\conttemp=50
\loop\ifnum\connttrre<\countref \advance\conttemp by 1
\advance\connttrre by 1
\item{\Reflab{\the\connttrre}}{\unhcopy\conttemp} \repeat}
\clearcountref \clearconnttrre
\catcode`@=12

\def\slashchar#1{\setbox0=\hbox{$#1$} \dimen0=\wd0
     \setbox1=\hbox{/} \dimen1=\wd1 \ifdim\dimen0>\dimen1
      \rlap{\hbox to \dimen0{\hfil/\hfil}} #1 \else
      \rlap{\hbox to \dimen1{\hfil$#1$\hfil}} / \fi}
\ifx\oldchi\undefined \let\oldchi=\chi
  \def\cchi{{\raise 1pt\hbox{$\oldchi$}}} \let\chi=\cchi \fi
\def\square{\hbox{{$\sqcup$}\llap{$\sqcap$}}}

\def\frac#1#2{{\textstyle{#1 \over #2}}}

\def\half{\ifinner {\scriptstyle {1 \over 2}}\else {1 \over 2} \fi}

\def\simge{\rlap{\raise 2pt \hbox{$>$}}{\lower 2pt \hbox{$\sim$}}}
\def\simle{\rlap{\raise 2pt \hbox{$<$}}{\lower 2pt \hbox{$\sim$}}}

\def\vbig#1#2{{\vbigd@men=#2\divide\vbigd@men by 2%
\hbox{$\left#1\vbox to \vbigd@men{}\right.\n@space$}}}

\null
%
%
%
%
%
\loadamsmath
\loadeuler
%
%
\nopagenumbers{\baselineskip=12pt
\line{\hfill OUTP-92-15P}
\line{\hfill CERN-TH-6594/92}
\line{\hfill hep-th/9208058}
\line{\hfill July, 1992}
\ifdoublepage \bjump\bjump\bjump\bjump\else\vfill\fi
\centerline{\capsone TAU--FUNCTIONS AND GENERALIZED}
\sjump\sjump
\centerline{\capsone INTEGRABLE HIERARCHIES}
\bjump\bjump
\centerline{\scaps Timothy Hollowood\footnote{$^1$}
{holl\%dionysos.thphys@prg.oxford.ac.uk\newline Address after
$1^{\rm st}$ October 1992: Theory Division, CERN, CH-1211 Geneva 23,
Switzerland}}
\sjump
\centerline{\sl Theoretical Physics, 1 Keble Road,}
\centerline{\sl Oxford, OX1 3NP, U.K.}
\sjump\sjump
\centerline{and}
\sjump\sjump
\centerline{\scaps J. Luis Miramontes\footnote{$^2$}{miramont@cernvm.cern.ch}}
\sjump
\centerline{\sl Theory Division, CERN,}
\centerline{\sl CH-1211 Geneva 23, Switzerland}
\vfill
\ifnum\unoduecol=2 \eject\null\vfill\fi
\centerline{\capsone ABSTRACT}
\sjump
\noindent
The tau-function formalism for a class of generalized ``zero-curvature''
integrable
hierarchies of partial differential equations, is constructed. The
class includes the Drinfel'd-Sokolov hierarchies. A direct relation
between the variables of the zero-curvature formalism and the
tau-functions is
established. The
formalism also clarifies the connection between the zero-curvature
hierarchies
and the Hirota-type hierarchies of Kac and Wakimoto.
\sjump
\ifnum\unoduecol=2 \vfill\fi
\eject}
\yespagenumbers\pageno=1

\def\gg{{\eufm g}}
\def\ss{{\eufm s}}
\def\s{{\bf s}}

\chapter{Introduction}

The evolution of the subject of integrable hierarchies of equations
has exhibited many unexpected twists. Arguably, the
first important mathematical result was the demonstration of
the integrability of the Korteweg-de Vries (KdV) equation
$$
{\partial u\over\partial t}={\partial^3 u\over\partial x^3}+
6u{\partial u\over\partial x}.
\nfr{kdveq}
Since then much effort has been devoted to finding the underlying
``causes'' for
integrability. Such an endeavour is intimately linked to the problem
of classification, because in a
general framework one can separate out the underlying important
``wheat'' of the problem from the example-dependent ``chaff''. We
believe that one of the most important and seminal works in this
regard was that of Drinfel'd and Sokolov [\Ref{DS}].
These authors provided the most general classification of
integrable hierarchies of equations up to that time. Their
construction is based on a zero-curvature, or Lax-type, method, where
integrability is manifest. The
central object in the construction are gauge fields in the
loop algebra of a finite Lie algebra. Crudely speaking, they arrive at
a picture where there is a {\it modified\/} KdV (mKdV)
hierarchy for each loop algebra, and then associated KdV hierarchies
for each of the nodes of the corresponding Dynkin diagram.

The Drinfel'd-Sokolov hierarchies make use of the ``principle''
gradation of the loop algebra in an essential way. In particular, the
construction involves the {\it principle Heisenberg subalgebra\/}. On
the other hand, it is well known that affine Kac-Moody algebras have
many Heisenberg subalgebras [\Ref{KACB},\Ref{KP}],
an observation that was exploited in [\Ref{GEN1}] (see
also [\Ref{GEN3}]) to
construct a more general class of integrable hierarchies. These
hierarchies share all the features of the Drinfel'd-Sokolov
hierarchies: there are mKdV and KdV-type hierarchies with
(bi-)Hamiltonian structures [\Ref{GEN2}].

However, there are ways, other than the zero-curvature method,
to investigate integrable hierarchies.
One of the most remarkable development in the subject started with
the work of R. Hirota (see for example [\Ref{HIR}]), who discovered a way to
construct various types of solutions to the hierarchies directly; in
particular the multiple soliton solutions can easily be found. This led to the
so-called ``tau-function'' approach pioneered by the Japanese school
(see for example [\Ref{JAP}]).
The idea is to find a new set of variables, called
the tau-functions, which then satisfy a new type of bi-linear
equation known as the Hirota equation. For instance, the tau-function
of the KdV equation -- in standard conventions -- is related
to the original variable by the celebrated formula
$$
u=2{\partial^2\over\partial x^2}\log\,\tau.
\nfr{kdvtau}
Correspondingly, for the {\it modified\/} KdV hierarchy the relation is
$$
\nu={\partial\over\partial x}\log\left({\tau_0\over\tau_1}\right),
\nfr{mkdvtau}
there being two separate tau-functions in this case.

Far from being just a new solution-method, the
tau-function approach uncovered a deep underlying structure of
integrable hierarchies. The story is quite long and complicated
involving unexpected connections to other branches of mathematics;
for which we refer the reader to the original literature
[\Ref{JAP},\Ref{SW},\Ref{WI}], and references therein.
It is clear from this approach that affine
Kac-Moody algebras (central extensions of loop algebras)
again play a central r\^ole. This was so in the
original work of [\Ref{JAP}], but made even clearer by Kac and
Wakimoto [\Ref{KW}]. In this latter work, the authors construct
hierarchies directly in Hirota form associated to vertex operator
representations of Kac-Moody algebras.

It certainly occurred to Kac and Wakimoto [\Ref{KW}] that there should be
a connection between their work and that of Drinfel'd and Sokolov:
both involving, as they do, Kac-Moody algebras. This present work is
an attempt to make this connection explicit. The situation for the
affine algebra $A_1^{(1)}$, which leads to the KdV and mKdV
hierarchies is well established [\Ref{SW},\Ref{WI}]. Some extensions to
other algebras and the {\it homogeneous\/} Heisenberg subalgebra were
considered in [\Ref{IM}]. Our approach follows very closely the spirit
of this latter work, and we shall only mention the {\it Grassmanian\/}
approach of the former work in passing.

The central goal of this work is to provide an explicit relation between the
tau-functions and the variables of the zero-curvature formalism. We
shall not find a one-to-one correspondence between the zero-curvature
hierarchies and the Kac-Wakimoto hierarchies: only a subset of both
classes are related.

The paper is organized as follows.
In section 2 we describe the general class of zero-curvature integrable
hierarchies of [\Ref{GEN1}], which contain the Drinfel'd-Sokolov
hierarchies as special cases.
Section 3 introduces the Kac-Wakimoto hierarchies which are defined
directly on the tau-functions, in terms of a vertex operator
representation of a Kac-Moody algebra.
Section 4 considers the ``dressing
transformation'' which allows one to construct solutions of the zero-curvature
hierarchies, and which also provides the key for establishing a
connection between the formalisms of sections 2 and 3. The explicit
connection is established in section 5 and some examples are considered
in section 6 for the purposes of illustration. Our conventions and some
properties of Kac-Moody algebras are presented in the appendix.

\chapter{The Zero-curvature Hierarchies}

The purpose of this section is two-fold: it should provide a
summary of some of the important details of refs.
[\Ref{GEN1},\Ref{GEN2}] and sets up some new results that
will be required in later sections. Our conventions concerning affine
Kac-Moody algebras are summarized in the appendix.

In refs. [\Ref{GEN1},\Ref{GEN2}], a
generalized integrable hierarchy was associated to each affine
Kac-Moody algebra $\gg$, a particular Heisenberg subalgebra
$\ss\subset\gg$ (with an associated gradation ${\bf s}'$)
and an additional gradation $\bf s$, such that ${\bf
s}'\succeq{\bf s}$, with respect to a partial ordering (see the appendix). The
auxiliary gradation $\s$ sets the ``degree of modification'' of the
hierarchy: the larger $\s$ becomes the more ``modified'' the hierarchy becomes.
In [\Ref{GEN1}], the construction was undertaken in the loop algebra
(Kac-Moody algebra with zero centre),
whereas for present purposes, it will actually prove more
convenient to present the construction in a representation
independent way in the full Kac-Moody algebra with centre; although we should
stress that the resulting hierarchy of equations is identical.

There is a flow of the hierarchy for each element of $\ss$ of
non-negative ${\bf
s}'$-grade, this is the set $\{b_j,\ j\in E\geq0\}$. The
flows are defined in terms of the gauge connections, or ``Lax
operators'', of the form
$$
{\cal L}_j={\partial\over\partial t_j}-b_j-q(j)\qquad j\in E\geq0,
\efr
where $q(j)$ is a function of the $t_j$'s on the
intersection
$$
Q(j)\equiv\gg_{\geq0}({\bf s})\bigcap\gg_{<j}({\bf s}').
\nfr{spac}
In order to ensure that the flows $t_j$ are uniquely associated to
elements of the set $\{b_j,\ j\in E\geq0\}$ we will also, without loss
of generality, demand that $q(j)$ has no constant terms proportional to
$b_i$ with $i<j$. The integrable hierarchy of equations is defined by
the zero-curvature conditions
$$
[{\cal L}_i,{\cal L}_j]=0.
\nfr{hier}

In general, the above systems exhibit a {\it gauge invariance\/} of the
form
$${\cal L}_j\mapsto U{\cal L}_jU^{-1}
\nfr{gaug}
preserving $q(j)\in Q(j)$,
where $U$ is a function on the group generated by the finite
dimensional subalgebra given by the intersection
$$
P\equiv\gg_0({\bf s})\bigcap\gg_{<0}({\bf s}').
\nfr{gaugespace}
The equations of the hierarchy are to be thought of as equations on
the equivalence of classes of $Q(j)$ under the gauge
transformations. Notice that if ${\bf s}\simeq{\bf s}'$ then
$P=\emptyset$. The only difference between the situation in
[\Ref{GEN1}] and the situation here, is that $q(j)$ may have a component
in the center of $\gg$, say $q_c(j)$. This also means that
the system is also
gauge invariant under \gaug\ with $U$ being just a function, i.e. related
to the exponentiation of the center of $\gg$. Obviously, only $q_c(j)$ is
sensitive to these particular transformations and
they can, in fact, be used to set $q_c(j)$ to any arbitrary
value. Consequently, we conclude that it is not
a dynamical degree of freedom but a purely gauge dependent quantity.
This, and the fact that $q_c(j)$ cannot
contribute to the time evolution of the other components of $q(j)$,
is the reason why the resulting hierarchy is identical to the one
constructed in [\Ref{GEN1}].

The equations of the hierarchy \hier\ can be interpreted as a
system of partial differential equations on some set of functions in a
number of different ways. For each {\it regular\/} element $b_k\in\ss$
($k>0$), so that $\gg$ admits the decomposition
$$
\gg=\ss\oplus{\rm Im}({\rm ad}\,b_k),
\nfr{decom}
(we shall use the notation $\ss^\perp$ to
denote the complement of $\ss$, meaning $\gg=\ss\oplus\ss^\perp$)
we may regard \hier\ as an integrable hierarchy of
partial differential equations on the functions $q(k)$, modulo the
action of the gauge symmetry discussed above. (In the language of
[\Ref{GEN1}] these are the ``type-I'' hierarchies of equations.)

Below we repeat some of the analysis of [\Ref{GEN1}], to show how the
results of that reference are modified when the algebra has a
non-trivial centre. First of all, we consider the analogue of
Proposition 3.2 of [\Ref{GEN1}].

\sjump\sjump
\noindent
{\bf Proposition 2.1}
{\it For a given $b_k\in\ss$, for which $\gg$ has the decomposition \decom,
there is a unique $y\in\ss^\perp_{<0}(\s')$ and
$h(k)\in\ss_{<k}(\s')$, which are functions of $q(k)$ and its $t_k$
derivatives, such that
$$
q(k)=-\Phi\left({\partial\over\partial t_k}-b_k-h(k)\right)
\Phi^{-1}-b_k,
\nfr{pro}
where $\Phi=\exp\,y$.}

\sjump
\noindent
{\it Proof\/}. The proof is exactly the same as that of Proposition
3.2 of [\Ref{GEN1}], the only difference being that now $q(k)$ has a
component in the centre of $\gg$. We equate
terms in \pro\ of equal $\s'$-grade to get a recursion relation of the form
$$
h_j(k)+[b_k,y_{j-k}]=\star.
\nfr{rec}
In the above $h_j(k)$ and $y_j$ are the components of $h(k)$
and $y$ of $\s'$-grade equal to $j$, and $\star$ denotes terms which
depend on $h_i(k)$, for $i>j$, and $y_i$, for $i>j-k$, and $q(k)$. The
proof proceeds by induction. The first equation of the series states
that
$$
h_{k-1}(k)+[b_k,y_{-1}]=q_{k-1}(k).
\nfr{blob}
We now appeal to the decomposition \decom\ in order to solve uniquely for
$h_{k-1}(k)$ and for $y_{-1}$. The same decomposition means that we can
solve \rec\ iteratively for $y$ and $h(k)$. $\square$

\sjump\sjump
Once more, the only difference between the situation in [\Ref{GEN1}] and
the situation here, is that both $h(k)$ and $q(k)$ have a component in
the centre of $\gg$, say $h_c(k)$ and $q_c(k)$, respectively, and one
finds $q_c(k)=h_c(k) +\left(\Phi b_k\Phi^{-1}\right)_c$.

The arguments of [\Ref{GEN1}] can then be applied, with some minor
modifications, to find the
other variables $q(j)$, for $j\neq k$, in terms
of $q(k)$ and its $t_k$ derivatives:
$$
q(j)=P_{\geq0[\s]}\left(\Phi b_j\Phi^{-1}\right)-b_j + h_c(j),
\nfr{othpot}
where $P_{\geq0[\s]}$ is the projector onto $\gg_{\geq0}(\s)$.
The variables $q(k)$ then satisfy the partial differential equations
$$
{\partial q(k)\over\partial t_j}=\left[q(j)+b_j,{\cal L}_k\right].
\nfr{hic}

Before proceeding, let us clear up some technical details. Proposition
3.6 of ref. [\Ref{GEN1}] states that the quantities $h(k)$ proportional
to elements of $\ss$ with $\s$-grade $\geq0$ are constants under the
flows \hic. Hence, they
contribute constant terms to $q(k)$ proportional to elements of the
Heisenberg subalgebra $b_j$ with $j<k$; an eventuality that we
disallowed in the discussion following equation \spac. Hence we
must impose the conditions $P_{\geq0[\s]}[h(k) - h_c(k)]=0$.
In contrast, the element of $h(k)$ in the center cannot be set to
zero: it has to be compatible with the zero curvature conditions \hier\
$$
\eqalign{
&{\partial h_{-m}(k)\over\partial t_j} -{\partial
h_{-m}(j)\over\partial t_k} = 0\cr
&{\partial h_c(k)\over\partial t_j} -
{\partial h_c(j)\over\partial t_k} = c \left( j h_{-j}(k)
- k h_{-k} (j) \right) \qquad j,k,m\in E\geq0\,.\cr}
\nfr{centhier}
We notice that the value of $h_c(k)$ is
completely arbitrary, up to these consistency equations.

The equations of the
hierarchy are to be thought of a set of partial differential
equations on a consistent gauge slice of $q(k)$, denoted $\tilde
q(k)$, under the gauge symmetry \gaug. The
hierarchy of equations, which are labelled by the data
$\{\gg,\ss,\s;b_k\}$, are then of the form
$$
{\partial\tilde q(k)\over\partial t_j}=F_j\left(\tilde
q(k),{\partial\tilde q(k)\over\partial t_k},\ldots\right)\qquad j\in
E\geq0,
\nfr{eqhier}
for some functions $F_j$ of $\tilde q(k)$ and its $t_k$-derivatives.

Notice that there is one-to-one correspondence between the solutions
of two hierarchies $\{\gg,\ss,\s;b_k\}$ and $\{\gg,\ss,\s;b_l\}$ (where
$b_k$ and $b_l$ both admit the decomposition \decom). The maps are
given by \othpot. In this sense, one does not distinguish between such
hierarchies. However, it was further shown in [\Ref{GEN2}], that the
above system of
equations could be written in a one-parameter family of coordinated
Hamiltonian forms. In particular, one of the Poisson bracket algebras
is a classical $W$-algebra. So there exists a classical $W$-algebra
associated to each $\{\gg,\ss,\s;b_k\}$ hierarchy. Although the
hierarchies corresponding to different $b_k$'s are in a sense the
same as regards their space of solutions,
the associated canonical formalisms are not the same and
different $W$-algebras are obtained.
For instance the hierarchies constructed by Drinfel'd and Sokolov, with
$\gg=sl(n)^{(1)}$, $\ss$ being the {\it principle\/} Heisenberg
subalgebra and $\s=(1,0,\ldots,0)$ (the homogeneous gradation)
lead to the $W_n^{(k)}$-algebras, where $k$
labels the different choices for the element $b_k\in\ss$.

We now return to the question of gauge invariance.
A convenient choice for the gauge slice is suggested by the following
proposition.

\sjump\sjump
\noindent
{\bf Proposition 2.2} {\it There exists a consistent gauge slice where
$$
P_{0[\s]}(y)=0,
\efr
where $y$ from Proposition 2.1 is considered as a function
of $q(k)$ and its $t_k$ derivatives.}

\sjump
\noindent
{\it Proof\/}. Gauge transformations act on $y$ as
$\Phi'=U\Phi$, with $U=\exp\,u$ where $u$ is a function on the algebra $P$ in
\gaugespace. Denoting the components of $y'$, $y$ and $u$ with zero
$\s$-grade as $y'_0$, $y_0$ and $u_0$, respectively, projecting onto
zero $\s$-grade we have
$$
\exp y_0'=\exp u_0\exp y_0.
\efr
Therefore, by choosing $u_0=-y_0$ we can gauge away the component
$P_{0[\s]}(y)$. Notice that this is consistent with
$y\in\ss_{<0}^\perp(\s')$ because $P\cap\ss=\emptyset$. $\square$

\sjump\sjump
The result of this proposition is that there exists a unique gauge
slice $\tilde q$ for which $y$ ($=\tilde y$) is a function on
$\ss^\perp_{<0}(\s)$;
from now on we will assume that this gauge slice has been chosen and
we shall denote $\exp\,\tilde y\equiv\tilde\Phi$.
As we have noted, if ${\bf s}\simeq{\bf
s}'$, then there is no gauge symmetry in the hierarchy leading to a
{\it modified\/} hierarchy (mKdV hierarchy) in the language of
[\Ref{DS}]. For ${\bf s}\prec{\bf s}'$ the
hierarchies are {\it partially modified\/} (pmKdV hierarchies); these
include the KdV hierarchies for which $\bf s$ is a ``minimal''
gradation, {\it i.e.\/} one for all the $s_j$ are equal to zero except
for say $s_k=1$.

For a given choice of Heisenberg subalgebra $\ss$,
the pmKdV hierarchies are related to the
mKdV hierarchy by a {\it Miura map\/} which takes solutions of the
mKdV hierarchy into solutions of the pmKdV hierarchy. The Miura maps
have been discussed in detail in [\Ref{GEN2}], however in the present
context they can be discussed in a slightly different way. Given two
hierarchies $\{\gg,\ss,\s_1;b_k\}$ and $\{\gg,\ss,\s_2;b_k\}$, with
$\s_2\succeq\s_1$, it follows that a solution of the second hierarchy
gives a solution to the first hierarchy since $Q(k;\s_2)\subset
Q(k;\s_1)$ (where we have made the dependence of the space $Q(k)$ on
the ``degree of modification'' explicit). With the choice of gauge in
Proposition 1.2 we can make the Miura map more explicit.

\noindent
\sjump\sjump
{\bf Proposition 2.3} {\it The Miura map, which takes solutions of a hierarchy
$\{\gg,\ss,\s_2;b_{k}\}$ to a solution of the hierarchy
$\{\gg,\ss,\s_1;b_{k}\}$ with $\s_2\succeq\s_1$, with the choice of
gauge in Proposition 2.2, is the projection
$$
\tilde y_1=P_{<0[\s_1]}(\tilde y_2),
\nfr{koo}
where $\tilde y_2$ is considered as a function of $\tilde q_2(k)$ and
$\tilde q_1(k)$ is then given in terms of $\tilde q_2(k)$ via
\koo\ and \pro.}

\sjump
\noindent
{\it Proof\/}. The existence of the Miura map follows from the fact
that if $\s_2\succeq\s_1$ then $Q(k;\s_2)\subset Q(k;\s_1)$, hence
$\tilde q_2(k)\in Q(k;\s_1)$. In order to ensure that the imagine
under the Miura map is in the gauge of Proposition 2.2 one has to
``gauge away'' $P_{0[\s_1]}(\tilde y_2)$, and hence the Miura map can
be thought of as the projection in \koo. $\square$

\sjump\sjump
Along with each integrable hierarchy there is an associated linear
problem:
$$
\left({\partial\over\partial t_j}-b_j-q(j)\right)\Psi=0\qquad j\in E\geq0,
\nfr{linear}
where $\Psi$ is a function of the $t_j$'s on the group $G$ formed by
{\it exponentiating\/} $\gg$.

We now prove a central theorem.

\sjump\sjump
\noindent
{\bf Theorem 2.4} {\it There is a one-to-one
map from solutions of the (gauge fixed) associated linear problem of
the form
$$
\tilde\Psi=\Theta\Gamma,
\nfr{hz}
where
$$
\Gamma=\exp\left[\sum_{j\in E\geq0}t_jb_j \right],
\nfr{defgam}
and $\Theta$ being a function on the subgroup $U_-(\s)$, and solutions
of the hierarchy \eqhier\ with the arbitrary functions $h_c(k)$ fixed
by the conditions $\partial h_c(k)/\partial t_j + c\,kh_{-k}(j)=0$ for
any $j,k\in E\geq0$.}

\sjump
\noindent
{\it Proof\/}.
First of all, using \pro\ and the equations for $h_c(k)$, the  gauge
fixed Lax operators can be written in the form
$$ {\cal L}_j = {\partial\over\partial t_j} + \left(\tilde\Phi
{\rm e}^\omega\right)
\left({\partial\over\partial t_j} - b_j\right)
\left(\tilde\Phi{\rm e}^\omega \right)^{-1}, \nfr{diagon}
where
$$
\omega =  \int^{t_m}\, dt_{m}'\,[h(m) - h_c(m)] \in \ss_{<0[\s']} ,
\nfr{integral}
with the integral defined along the $t_m$-flow. Notice that $\omega$
is a well defined function independently of the choice of $m$
because $h(j)$ satisfies \centhier. Therefore, we can build a solution
of the linear problem of the form \hz\ with
$$
\Theta = \tilde\Phi {\rm e}^\omega
\nfr{soldir}

On the other hand, if we are given a solution of the linear
problem of the form \hz\ then it is straightforward to see that
$$
\tilde q(k)=-\Theta\left({\partial\over\partial
t_k}-b_k \right)\Theta^{-1}-b_k,
\nfr{sollin}
however, what is not so clear is that the quantity
$P_{\geq0[\s]}\left[ h(k)-h_c(k)\right]$
equals zero for $\tilde q(k)$ given by
\sollin. To see this one first finds $\tilde\Phi$ as a function of
$\tilde q(k)$ and its $t_k$ derivatives, via Proposition 2.1, and
therefore ultimately as a function of $\Theta$ via \sollin.
The result is that
$\tilde\Phi^{-1}\Theta= {\rm e}^ u$ with $u=\sum_{j<0} u_j b_j \in
\ss_{<0[\s']}$, which gives
$$
\eqalign{
h(k) &= -(\Theta^{-1}\tilde\Phi)^{-1}{\partial\Theta^{-1}\tilde\Phi
\over\partial t_k}- b_k + \left[\left(   \Theta^{-1}\tilde\Phi
\right)^{-1}
b_k \Theta^{-1} \tilde\Phi\right]_c  \cr
&= \sum_{j<0} {\partial u_j\over\partial t_k} b_j - cku_{-k}\cr}
\efr
Therefore, it satisfies the two conditions
$P_{\geq0[\s]}\left[h(k)- h_c(k)\right] =0 $, and ${\partial h_c(k)
/ \partial t_j} + c\,k h_{-k} (j)=0$.
$\square$

\sjump\sjump

The result \sollin\ admits a very useful simplification. The solution
$\tilde q(k)$ has $\s$-grade $\geq0$, hence using the fact that
$\Theta\in U_-(\s)$
$$
\tilde q(k)=P_{\geq0[\s]}\left(\Theta b_k\Theta^{-1}\right)-b_k.
\nfr{result}
The importance of this result is that it only needs the finite number
of terms of $\Theta$ with $\s$-grade greater than $-k-1$ to be applied.

\chapter{The Kac-Wakimoto Hierarchies and Tau-Functions}

In this section we provide a short review of the construction of
integrable hierarchies by Kac and Wakimoto [\Ref{KW}]. This will lead
us to introduce the tau-functions.

The construction of Kac and Wakimoto leads directly to the
equations of the hierarchy in {\it Hirota form\/} [\Ref{HIR}]. The idea is the
following: the tau-function $\tau_{\bf s}$ associated to an
integrable highest weight representation $L({\bf s})$ of an
affine Kac-Moody algebra $\gg$ is characterized by saying that it lies
in the $G$-orbit of the highest weight vector $v_{\bf s}$. Here $G$ is
the group associated to $\gg$.

Let $\{u_i\}$ and $\{u^j\}$ be dual
bases of the larger algebra $\gg\oplus{\Bbb C}d$,
with respect to the non-degenerate bi-linear inner product
$(\cdot|\cdot)$. It can be shown [\Ref{KW},\Ref{PK}]
that $\tau_{\bf s}$ lies in
the $G$-orbit of $v_{\bf s}$ if and only if
$$
\sum u_j\otimes u^j\left(\tau_{\bf s}\otimes\tau_{\bf
s}\right)=(\Lambda_{\bf s}|\Lambda_{\bf s})\tau_{\bf s}\otimes\tau_{\bf s},
\nfr{hirota}
where $\Lambda_{\bf s}$ is the eigenvalue of $\gg_0({\bf s})$ on
$v_{\bf s}$. We can think of ${\cal C}=\sum u^j\otimes u_j$ as a generalized
Casimir operator. Furthermore, the condition \hirota\ is also
equivalent to the statement that
$$
\tau_{\bf s}\otimes\tau_{\bf s}\in L(2{\bf s}).
\nfr{highcond}

It follows from the definition of the action of a group on a tensor
product that, for the representation $L(s)$,
$$
\tau_{\s}=\bigotimes_{i=0}^r\left\{\tau_i^{\otimes\,s_i}\right\},
\nfr{fact}
where $\tau_i$ is the tau-function corresponding to the fundamental
representation with $s_j=\delta_{ij}$.

At the moment, the conditions equations \hirota\ are completely
``group theoretic'', with no apparent connection to integrable hierarchies of
equations. However, for cases where the representations are of ``vertex
type'', so they are carried by Fock spaces, then \hirota\ can be interpreted as
differential equations on the tau-functions. In fact, they are
precisely the Hirota equations of an integrable hierarchy.
In order to explain this, we restrict ourselves to cases where
$\gg=g^{(1)}$ is the
untwisted affinization of a finite simply-laced algebra $g$. In that
case, level one representations (or {\it basic\/} representations,
those for which $s_j=\delta_{ji}$
for some $i$ such that $k_{i}^{\vee} = k_i=1$) are
isomorphic to the Fock space of
any one of the Heisenberg subalgebras of $\gg$. It is known that
inequivalent Heisenberg subalgebras are classified by the conjugacy
classes of the Weyl group of $g$ [\Ref{KACB},\Ref{KP}].
The connection between the Weyl
group element, say $w$ (up to conjugacy), and the associated Heisenberg
subalgebra $\s_w$, is that there is a lift of $w$, denoted $\hat w$, onto
$\gg$, which acts on the Heisenberg subalgebra as
$$
\hat w\left(b_j\right)=\exp\left({2\pi ij\over N}\right)b_j.
\efr

The Heisenberg subalgebra $\ss_w$ is realized on the Fock space ${\Bbb
C}[x_j,\ j\in E>0]$ in the standard way: $c=1$ and
$$
b_j=\cases{{\partial\over\partial x_j}\quad&$j>0$\cr
-jx_{-j} &$j<0$.\cr}
\efr
A rather different treatment is required for any zero-graded
generators of $\ss_w$ which correspond to the invariant subspace of
$w$. These zero-modes are represented on the space
$$
{\Bbb C}(Q)=\{\exp\,\beta\cdot x_0,\quad \beta\in Q\},
\efr
where $Q$ is the root lattice of $g$ projected onto the invariant
subspace of $w$. $b_0$ acts as $\partial/\partial x_0$ so
$$
b_0e^{\beta\cdot x_0}=\beta e^{\beta\cdot x_0},\qquad \beta\in Q.
\efr

The level-one representation is isomorphic to ${\Bbb
C}[x_j]\otimes{\cal V}$,
where ${\cal V}={\Bbb C}(Q)\otimes V$ is the
{\it zero-mode space\/}. Here, $V$ is an additional
finite-dimensional vector space [\Ref{KP},\Ref{VERTEX}].
The elements of $\gg$ not in $\ss_w$ are the modes of vertex
operators, the centre is the identity ($c=1$), and
the derivation $d_{\bf s'}$ is the zero-mode of the Sugawara
current, up to a constant. Notice that
the construction does not distinguish between the different level-one
representations of $g^{(1)}$, this is a reflection of the fact that
all such representations are isomorphic due to symmetries of the
extended Dynkin diagram.

The equations \hirota\ after expressing the generators of $\gg$ in terms
of operators on the Fock-space, are then bi-linear Hirota equations
for the functions $\tau_i^{(\beta)}(x_j)$, which are projections onto a
basis for ${\Bbb C}(Q)$:
$$
\tau_i(x_0;x_j)=\sum_{\beta\in Q}\tau_i^{(\beta)}(x_j)e^{\beta\cdot x_0}.
\nfr{compon}

We wish to emphasize that there is a different realization of each
level-one representation for each inequivalent Heisenberg subalgebra
of $\gg$, and moreover, although these realizations are isomorphic as
representations they lead to different Hirota equations for the
corresponding tau-functions.

In the following we shall often
deal with the vertex representation of $L(\s)$ realized on the tensor product
of fundamental representations, where $s_i$ gives the multiplicity
of the $i^{\rm th}$ fundamental representation
in the product (so any non-zero $s_i$ corresponds to $k_{i}^{\vee}=
k_i=1$).
They will be carried by a tensor product of the Fock spaces:
$$
\bigotimes_{i=1}^N\left\{{\Bbb C}[x_j,\ j\in E>0]\otimes{\cal
V}\right\},
\efr
where $N=\sum_{i=0}^rs_i$. We shall use the notation $x_j^{(i)}$ to
indicate the Fock space variables of the $i^{\rm th}$ space in the
tensor product, and $x_j\equiv\sum_{i=1}^Nx_j^{(i)}$.

\chapter{Dressing Transformations}

In this section we define a set of transformations on the
zero-curvature hierarchies,
which take known solutions of a hierarchy to new solutions. These
``dressing transformations'' will be crucial for establishing the link
to the tau-function formalism of the next section.

Consider the gauge-fixed linear problem associated to the hierarchy:
$$
\left({\partial\over\partial t_j}-b_j-\tilde q(j)\right)\tilde\Psi=0,
\nfr{lin}
We have already established in Theorem 2.4
a map between solutions of the gauge-fixed linear problem of the form
$$
\tilde\Psi=\Theta\Gamma\qquad \Theta\in U_-(\s),
\nfr{form}
where $\Gamma$ is defined in \defgam, and solutions of the gauge-fixed
hierarchy.

The {\it dressing transformation\/} is a map between solutions of \lin,
which preserves the form \form. Many of the technical aspects of these
transformations are considered in [\Ref{WI}] for the case
$\gg=A_1^{(1)}$ and the principle Heisenberg subalgebra; however,
our approach is closer to that of ref.
[\Ref{IM}]. One difference with these other works is that we will deal
with the KdV hierarchies directly rather than considering the mKdV
hierarchies and then using the Miura map to find the resulting
transformations for the KdV hierarchies.

\sjump\sjump
\noindent
{\bf Theorem 4.1} {\it Given a solution $\tilde\Psi_1$ of the linear
problem \lin, of the form \form, and $g\in
G$, with $\tilde\Psi_1\cdot g=(\tilde\Psi_1\cdot
g)_-(\tilde\Psi_1\cdot g)_0
(\tilde\Psi_1\cdot g)_+$ lying in the
``big cell'' $U_-(\s)H(\s)U_+(\s)$ then
$$\tilde\Psi_2={\cal D}_g\left(\tilde\Psi_1\right)
\equiv\left[\left(\tilde\Psi_1\cdot g\right)_-\right]^{-1}\tilde\Psi_1=
\left(\tilde\Psi_1\cdot g\right)_0\left(\tilde\Psi_1\cdot g\right)_
+g^{-1},
\nfr{dress}
is also a solution of the linear problem with the form \form.}

\sjump
\noindent
{\it Proof\/}. We have to show that $\tilde q(j)$ defined by
$$
{\partial\over\partial t_j}\tilde\Psi_2=\left(\tilde
q(j)+b_j\right)\tilde\Psi_2,
\efr
lies in the subspace $Q(j)$ defined in \spac.  The proof follows from the
two expressions for $\tilde\Psi_2$ in \dress. From the first
expression we find
$\tilde q(j)+b_j=b_j+$ terms with $\s'$-grade $<j$ and from the second
expression we find $\tilde q(j)+b_j=$ terms with $\s$-grade $\geq0$.
Hence $\tilde q(j)$ lies in $Q(j)$ as required. Moreover,
it is clear from the first expression in \dress\ that $\tilde\Psi_2$ has
the form of \form. $\square$

\sjump\sjump
\noindent
{\bf Corollary}. {\it The expressions
$$
\tilde\Psi={\cal D}_g\left({\Gamma}\right),
\efr
where $\Gamma$ is defined in \defgam, are solutions of the linear
problem of the form \form\
and consequently using Theorem 2.4 solutions of the hierarchy.}

\sjump
\noindent
{\it Proof\/}. The proof is elementary. One just has to notice that
$$
\Gamma=\exp\left[\sum_{j\in E\geq0}b_jt_j\right],
\efr
satisfies the linear problem with $\tilde q(j)=0$, and obviously has
the form \form. $\square$

\chapter{The Zero-Curvature hierarchies and the Tau-Functions}

In section 2 we introduced a series of integrable hierarchies
constructed via a zero-curvature method, whilst in section 3
we described a series of hierarchies in the form of a set of Hirota
equations. It is now time to connect these two formalisms, the bridge
being provided by the dressing transformation.

Let us review again the situation in the two formalisms. In the zero
curvature formalism a hierarchy was defined in terms of the following
data $\{\gg,\ss,\s;b_k\}$:

\sjump\sjump
\noindent
(i) An affine Kac-Moody algebra $\gg$.

\noindent
(ii) A Heisenberg subalgebra $\ss\subset\gg$ (with an associated
gradation $\s'$).

\noindent
(iii) A gradation $\s$ (``the degree of modification'') such that
$\s\preceq\s'$.

\noindent
(iv) An element $b_k\in\ss$ with positive grade such that $\gg$
admits the decomposition \decom, {\it i.e.\/} ${\rm Im}({\rm
ad}\,b_k)=\ss^\perp$.

\sjump\sjump
In the tau-function formalism, a hierarchy of Hirota equations was
associated to the following data $\{g,w,\s\}$:

\sjump\sjump
\noindent
(i) A simply-laced finite Lie algebra $g$.

\noindent
(ii) A vertex operator realization of $L(\s)$, associated to the
untwisted affinization of $g$, corresponding to some
conjugacy class of the Weyl group of $g$, containing $w$. This
requires that $s_i=0$ if $k_i>1$.

\sjump\sjump
To connect the zero-curvature and tau-function formalisms we notice that the
dressing transformations introduced in the last section involve a
group element $g\in G$. The idea is to use the group element which appears
in the characterization of the tau-function to ``dress'' the ``vacuum''
solution to the linear system [\Ref{IM}], in the manner of the corollary of
Theorem 4.1. We are led to the following key theorem.

\sjump\sjump
\noindent
{\bf Theorem 5.1}. {\it There exists a map from solutions of the
Kac-Wakimoto hierarchy associated to the data $\{g,w,\s\}$ (with the gradation
associated to the Heisenberg subalgebra $\ss_w$ satisfying $\s_w\succeq\s$
(and also $s_i>0$ only if $k_i=1$) and a zero-curvature hierarchy
associated to the data $\{g^{(1)},\ss_w,\s;b_k\}$, given by
$$
\Theta^{-1}\cdot v_\s={\tau_\s(x_j+t_j)\over\tau_\s^{(0)}(t_j)},
\efr
where $\Theta\in U_{-}(\s)$ gives $q(k)$ via \result}

\sjump
\noindent
{\it Proof\/}. Consider solutions of the linear problem which
follow from the corollary of Theorem 4.1, from which we deduce that in
representation $L(\s)$
$$
\Gamma\cdot\tilde\Psi^{-1}\cdot v_\s=\left(\Gamma\cdot g\right)_-\cdot v_\s.
\efr
Now we have
$$\tau_\s(x_j+t_j)=
\Gamma\cdot g\cdot v_\s=(\Gamma\cdot g)_-(\Gamma\cdot g)_0\cdot v_\s
=f(t_j)(\Gamma\cdot g)_-\cdot v_\s,
\efr
where $f(t_j)$ is the eigenvalue of $(\Gamma\cdot g)_0$ on $v_\s$. (It
should be emphasized that the representation is defined by the Fock
space of the variables $x_j$; the variables $t_j$ are to be thought of
as auxiliary variables.) To compute the eigenvalue we notice that
$$
\tau_\s(x_j+t_j)=\Gamma\cdot g\cdot v_\s=(\Gamma\cdot g)_0\cdot v_\s+\cdots,
\efr
where the ellipsis represents states in the representation with lower
$\s$-grade; in other words $f(t_j)$ is given by the coefficient of the
projection of $\tau_\s(x_j+t_j)$ onto $v_\s$; but this is
$$
f(t_j)=\tau_\s^{(0)}(t_j),
\efr
where $\tau_\s^{(0)}$ is the component of the tau-function with zero
``momentum'' in \compon\note{We remark at this point that
$\tau_\s^{(0)}(t_j)$ is a scalar quantity, because if
$\s_w\succeq\s$ then the auxiliary vector space $V$ is trivial
(dim$V$=1), which follows from the corollary to Lemma A.1}. Hence
$$
\Theta^{-1}\cdot v_\s=
\Gamma\cdot\tilde\Psi^{-1}\cdot v_\s={\tau_\s(x_j+t_j)\over\tau_\s^{(0)}(t_j)},
\nfr{geto}
where $\tau_\s=g\cdot v_\s$ is the tau-function for the representation
$L(\s)$. Now given $\tau_\s$ then \geto\ uniquely determines $\Theta\in
U_-(\s)$ -- since $U_-(\s)$ is faithful on $v_\s$ --
 and hence via Theorem 2.4 a solution of the zero-curvature hierarchy
$\{g^{(1)},\ss_w,\s;b_k\}$, for any $b_k$ admitting the decomposition
\decom. $\square$

\sjump\sjump
The theorem allows one to find the direct relation between the
tau-functions and the variables of the zero-curvature hierarchies.
Notice that not all zero-curvature hierarchies can be related to
tau-functions ($\s$ must correspond to products of level one
representations), and conversely not all Kac-Wakimoto hierarchies can
be related to zero-curvature hierarchies (due to the condition
$\s'\succeq\s$). Notice that in our formalism the KdV (and pmKdV)
hierarchies can be dealt with directly without recourse to the Miura
map, indeed, one of the pleasant results of the above formalism is
that one can see immediately, from the ``degree of modification'' $\s$,
which tau-functions are required. The Miura map at the level of the
tau-functions is the trivial statement, which follows from \fact, that
if $\s_2\succeq\s_1$ then $\tau_{\s_1}\subset\tau_{\s_2}$.

\chapter{Examples}

In this section we consider in some detail some examples of the
preceeding formalism, in order to illustrate some of the issues
involved. The main idea is to use Theorem 5.1 to find an expression
for the variables of the zero-curvature hierarchies in terms of the
tau-functions, generalizing the maps in \kdvtau\ and \mkdvtau. In
order to follow the calculations in this section some
knowledge of the vertex operator calculus is required, for which we
refer the reader to the original literature [\Ref{KW},\Ref{KP},\Ref{VERTEX}].

The Drinfel'd-Sokolov hierarchies [\Ref{DS}]
are recovered in our formalism by
choosing $\ss$ to be the principle Heisenberg subalgebra. For example,
consider the case when $\gg=sl(N)^{(1)}$.
The principle Heisenberg subalgebra has generators with grades in
$E=\{1,2,\ldots,N-1,\ {\rm mod}\,N\}$, and the associated gradation
is $\s'=(1,1,\ldots,1)$.
The basic representations of
$\gg=sl(N)^{(1)}$ are then represented in terms of the principle
Heisenberg subalgebra on the Fock space ${\Bbb C}[x_j,\ j\in E>0]$;
there are no zero-modes in this case and so ${\cal V}\simeq 1$.

The Drinfel'd-Sokolov hierarchies were originally defined in ref.
[\Ref{DS}] in terms of the
loop algebra $sl(N)^{(1)}$; taken to be
Laurent polynomials in a variable $z$ with coefficients in the
$N$-dimensional representation of $sl(N)$. The elements of the
principle Heisenberg subalgebra are
$$
b_j=\Lambda^j\qquad j\neq N{\Bbb Z},
\efr
where
$$
\Lambda=\pmatrix{&1&&&\cr &&1&&\cr &&&\ddots&\cr &&&&1\cr z&&&&\cr},
\efr
with zeros elsewhere.

The $sl(N)^{(1)}$ modified KdV hierarchy is generated from the Lax
operator
$$
{\cal L}_1={\partial\over\partial x}-\pmatrix{\nu_0&&&\cr &\nu_1&&\cr
&&\ddots&\cr &&&\nu_{N-1}\cr}-\Lambda,
\efr
where $x\equiv t_1$ and $\sum_{i=1}^N\nu_{i-1}=0$.
The form of this operator follows from the
systematic construction in \S1.
We now relate the variables $q(1)=\{\nu_i,\ i=0,1,\ldots,N-1\}$ to the
tau-functions $\tau_i$, $i=0,1,\ldots,N-1$. From \result\ we have
$$
q(1)=[\Theta_{-1},\Lambda],
\nfr{loo}
where $\Theta_{-1}$ is the component of $\Theta$ of $\s(=\s')$-grade
$-1$. Now $\Theta^{-1}\in U_-(\s)$, so we may write for some functions
$a_i$
$$
\Theta^{-1}=\exp\left[\sum_{i=0}^{N-1}a_ie^-_i+\cdots\right].
\efr
where the ellipsis represents terms with lower $\s$-grade.
Acting on the highest weight vector in the representation $L(\s)$ we have
$$
\Theta^{-1}\cdot
v_\s=\left[1+\sum_{i=0}^{N-1}a_ie^-_i+\cdots\right]\cdot v_\s.
\nfr{koj}
Using properties of the reducible representation
$L(\s)$ with $\s=(1,1,\ldots,1)$ in terms of vertex operators on the
Fock space $\otimes_{i=0}^{N-1}{\Bbb C}[x^{(i)}_j,\ j\in E>0]$ with
highest weight vector $v_\s=\bigotimes_{i=0}^{N-1} v_i$, one finds
$$
e_i^-\cdot v_j=\left(x^{(j)}_1\cdot v_j\right)\delta_{ij}.
\nfr{kij}
{}From \koj\ and \kij\  and using Theorem 5.1 one deduces
$$
a_i={\partial\over\partial x}\log\,\tau_i.
\efr
and therefore
$$
\Theta_{-1}=-\sum_{i=0}^{N-1}e_i^-{\partial\over\partial
x}\log\,\tau_i.
\efr
Now one can find $q(1)$ from \loo, however, to make the connection
with the formalism of Drinfel'd and Sokolov we must move to the loop
algebra. This is achieved by noting that in the loop algebra
$$e_i^-=\cases{{\bfmath e}_{i,i+1}\qquad&$i=1,2,\ldots,N-1$\cr
z^{-1}{\bfmath e}_{N,1}&$i=0$,\cr}
\efr
where ${\bfmath e}_{i,j}$ is the matrix with a 1 in the $(i,j)^{\rm
th}$ position and zero elsewhere. Applying \loo\ one
finds the well-known relation
$$
\nu_i={\partial\over\partial x}\log\left[{\tau_i\over\tau_{i+1}}\right],
\efr
with $\tau_{i+N}\equiv\tau_i$. The expression \mkdvtau\ for the
original mKdV hierarchy is a particular example of this.

There is no obstacle in extending the analysis to the
Drinfel'd-Sokolov KdV hierarchies, however, in the general case the
formulas are complicated and not very illuminating. Rather than
treating the general case we shall be satisfied with re-deriving the
famous relation \kdvtau\ of the original KdV hierarchy -- which arises
from choosing $\gg=sl(2)^{(1)}$ and $\ss$ to be the principle
Heisenberg subalgebra, as above. In this case $\s'=(1,1)$ (the
principle gradation) and
the ``degree of modification'' $\s=(1,0)$ (the homogeneous gradation).

The KdV hierarchy is defined via the Lax operator
$$
{\cal L}_1={\partial\over\partial x}-\pmatrix{w&0\cr v&-w\cr}-
\pmatrix{0&1\cr z&0\cr},
\efr
where as before $x\equiv t_1$. The gauge symmetry acts as
$$
{\cal L}_1\mapsto\pmatrix{1&0\cr g&1\cr}{\cal L}_1\pmatrix{1&0\cr
-g&1\cr}.
\efr
The choice of gauge made by Drinfel'd and Sokolov is
$$
\tilde q(1)=\pmatrix{0&0\cr -u&0\cr},
\nfr{can}
for which $u$ is then the conventional variable of the KdV hierarchy.

We now follow the same steps as for the mKdV hierarchies, but now
with $\s=(1,0)$. Putting
$$
\Theta^{-1}=\exp\left(ae_0^-+b[e^-_1,e^-_0]+\cdots\right),
\efr
for some functions $a$ and $b$,
where the ellipsis represents terms with lower $\s$-grade
which will not be required. Acting on the highest weight state one finds
$$
\Theta^{-1}\cdot
v_0=\left(1+ae^-_0+be^-_1e^-_0+\cdots\right)\cdot v_0.
\efr
Using properties of the vertex operator representation and
theorem 5.1 one finds
$$
a={1\over\tau}{\partial\tau\over\partial x},\qquad
b=-{1\over2\tau}{\partial^2\tau\over\partial x^2},
\efr
where $\tau\equiv\tau_0$ and $x\equiv t_1$. Moving to the loop algebra
using
$$e_0^-=\pmatrix{0&z^{-1}\cr 0&0\cr},\quad e_1^-=\pmatrix{0&0\cr
1&0\cr},
\efr
and evaluating \result\ we find
$$
\tilde q(1)=\pmatrix{-\tau'/\tau&0\cr -\tau''/\tau&\tau'/\tau\cr},
\efr
where $'\equiv{\partial/\partial x}$.
The result is not in the Drinfel'd and Sokolov gauge \can, however, it is
straightforward to find the gauge transformation connecting the gauge
choices. Making the required gauge
transformation one re-derives the classic result \kdvtau
$$
u=2{\partial^2\over\partial x^2}\log\,\tau.
\efr

As the last example we consider the ``homogeneous hierarchies'' which
are obtained by taking $\s=\s'=(1,0,0,\ldots,0)$, the homogeneous
gradation. This includes the non-linear Schr\"odinger hierarchy when
$\gg=sl(2)^{(1)}$. The homogeneous Heisenberg subalgebra is spanned,
in the loop algebra by $H\otimes z^n$, where $H$ is the Cartan
subalgebra of the finite horizontal subalgebra $g\subset\gg$. So at
each level there is a vector space of flows corresponding to the Cartan
subalgebra of $g$. The hierarchies are defined by the Lax operator
$$
{\cal L}_1={\partial\over\partial
x}-\sum_{\alpha\in\Phi_g}q^\alpha E_\alpha
-zH,
\efr
where $x\equiv t_1$ and the variables $q^\alpha$ are Cartan subalgebra-valued.
In the above we have introduced the Cartan Weyl basis for $g$ (see ref.
[\Ref{GEN1}] for a more thorough discussion), and $\Phi_g$ is the root
system of $g$.

Repeating the arguments above, we have for some arbitrary functions
$a^\alpha$ and $b$
$$\eqalign{
\Theta^{-1}\cdot v_\s&=\exp\left[\sum_{\alpha\in\Phi_g}a^\alpha
(E_\alpha)_{-1}+b\cdot H_{-1}+\cdots\right]\cdot v_0\cr
&=\left[1+\sum_{\alpha\in\Phi_g}a^\alpha
(E_\alpha)_{-1}+b\cdot H_{-1}+\cdots\right]\cdot v_0,\cr}
\efr
where the ellipsis represents terms of lower homogeneous grade, which
will not be required.
Now using properties of the vertex operator representation one finds
$$
a^\alpha={\tau^{(\alpha)}\over\tau^{(0)}}e^{\alpha\cdot t_0}
\quad \forall\alpha\in\Phi_g,\quad
b={\partial\over\partial x}\log\,\tau^{(0)}.
\efr
Moving to the loop algebra $(E_\alpha)_{-1}=z^{-1}E_\alpha$ and
$H_{-1}=z^{-1}H$ and applying \loo\ one finds
$$
q^{\alpha}=\alpha\,{\tau^{(\alpha)}\over\tau^{(0)}}\,
e^{\alpha\cdot t_0}\,.
\efr
This agrees with that found by Kac and Wakimoto [\Ref{KW}], and by
Imbens [\Ref{IM}], for the particular case when $\gg=sl(2)^{(1)}$
(which is the non-linear Schr\"odinger hierarchy) -- although in both
these references the trivial $t_0$ evolution is not considered.

\chapter{Discussion}

In this paper we have established the connection between
some of the zero-curvature hierarchies and some of the Kac-Wakimoto
hierarchies. We provided an algorithm for finding the variables of the
zero-curvature hierarchies in terms of the tau-functions.

Given the relation between the variables of the zero-curvature
hierarchies and the tau-functions it is
straightforward to write down the expression for
the multiple soliton solutions of the hierarchies.
If we denote by $V(\alpha,z)$ the vertex operator
associated to the orbit of the root $\alpha$ of the finite Lie
algebra, under the cyclic subgroup of the Weyl group of $g$ generated
by $w$ -- the Weyl group element that defines the vertex operator
representation -- then the $N$ soliton solution is given by the
tau-function
$$
\tau_\s(x_j)=(1+a_1V(\alpha_1,z_1))(1+a_2V(\alpha_2,z_2))\cdots
(1+a_NV(\alpha_N,z_N))\cdot v_\s,
\efr
where the $\alpha_p$'s are roots of $g$, indicating that each soliton
carries a ``flavour'' -- labelled by an orbit of a root under
the cyclic group generated by $w$ -- the $a_p$'s are constants and the
$z_p$'s are ``velocity parameters'' of the
solitons. Given some familiarity with vertex operator representations,
it is not difficult to find the explicit form for the soliton
solutions.

In our exposition we did not mention the {\it Grassmanian\/} approach
of refs. [\Ref{SW},\Ref{WI}], and so a few comments are in order.
Given that $\tau_\s=g\cdot v_\s$ for $g$
in the Group $G$ associated to the Kac-Moody algebra $\gg$, and the
fact that the highest weight vector $v_\s$ is annihilated by the
subalgebra generated by the $e_i^-$, for each $i$ such that $s_i=0$, and
the $e_i^+$, for $i=0,1,\ldots,r$,
two group element yield the same tau-function if they
correspond to the same class in the quotient $G/P_\s$, where $P_\s$ is
the {\it parabolic\/} subgroup generated by this subalgebra. Now consider
the case $\gg=sl(2)^{(1)}$. In this case solutions of the KdV
hierarchy are associated to $G/P$, where $P\equiv P_{(1,0)}$, and
solutions of the mKdV hierarchy are associated to $G/B$, where
$B=P_{(1,1)}$. In refs. [\Ref{SW},\Ref{WI}] it is shown that
$G/P$ is the Grassmanian and $G/B$ is a {\it flag manifold\/}, and
there exists a natural projection $G/B\rightarrow G/P$, which is
nothing else than the Miura map taking solutions of the mKdV hierarchy
to solutions of the KdV hierarchy. In our more general setting the
Grassmanian is replaced by $G/P_\s$, and there is a natural projection
$G/P_{\s_1}\rightarrow G/P_{\s_2}$, if $\s_1\succeq\s_2$, which is
again a geometrical statement of the Miura map between the
solutions of the two hierarchies.

One of the motivations which lies behind this work, comes from recent
developments regarding quantum gravity theories in two-dimensions.
In this context the partition function of pure gravity
is the tau-function of the KdV hierarchy, where the flows $t_i$ are the
coupling constants for all the operators in the theory (the
gravitational descendents of the puncture operator which couples to
$t_1\equiv x$), supplemented by an additional condition which is
called the ``string equation''. Remarkably, the string equation and
the condition that the partition function is a tau-function of the KdV
hierarchy are equivalent to the Virasoro constraints [\Ref{VIR}]:
$$
L_n\tau=0\qquad n\geq-1,
\efr
where the Virasoro generators are constructed from the Heisenberg
subalgebra. We shall show that there is a very natural generalization
of this structure to the other hierarchies that we have considered in
this paper; in the general case one obtains $W$-algebra constraints,
generalizing the situation for the $sl(N)$ Drinfel'd-Sokolov KdV
hierarchies [\Ref{VIR},\Ref{WCON}]
where the $W$-currents are constructed from the appropriate Heisenberg
subalgebra [\Ref{US}].

\acknowledgements
TJH would like to thank Merton College, Oxford, for
a Junior Research Fellowship. We would like to thank V.G. Kac for
making some useful comments about the manuscript which allowed us to
clarify certain points.

\appendix{}

In this appendix we review some of the details of affine Kac-Moody
algebras which will be important for the following constructions. A
complete treatment of such algebras may be found in [\Ref{KACB}], and
references therein.

An affine
Kac-Moody algebra $\gg$ is defined by a generalized Cartan matrix $a$, of
dimension $r+1$ and rank $r$, and is
generated by $\{h_i,e_i^+,e_i^-,\ i=0,1,\ldots,r\}$ subject to the
relations
$$\matrix{
[h_i,h_j]=0,\quad &[h_i,e^\pm_j]=\pm a_{ij}e^\pm_j,\cr
[e_i^+,e_j^-]=\delta_{ij}h_i,\quad
&\left({\rm ad}\,{e^\pm_i}\right)^{1-a_{ij}}\left(e_j^\pm\right)=0.\cr}
\efr
The algebra $\gg$ has a centre ${\Bbb C}c$ where
$$
c=\sum_{i=0}^rk_i^\vee h_i,
\efr
where $k_i^\vee$ (the {\it dual Kac labels\/}) are the components of
the left null eigenvector of $a$:
$$
\sum_{i=0}^rk_i^\vee a_{ij}=0.
\efr

A derivation $d_{\bf s}$, with ${\bf s}=(s_0,s_1,\ldots,s_r)$
being a set of
$r+1$ non-negative integers [\Ref{KACB}],
induces a ${\bf Z}$ grading on $\gg$ which we label $\bf s$:
$$
[d_{\bf s},e_i^\pm]=\pm s_ie_i^\pm,\qquad [d_{\bf s},h_i]=0.
\efr
Under ${\bf s}$, $\gg$ has the eigenspace decomposition
$$
\gg=\bigoplus_{j\in{\Bbb Z}}\gg_j({\bf s}).
\efr
We shall often use the notation like $\gg_{>k}(\s)=\bigoplus_{i>k}
\gg_i(\s)$.
There exists a partial ordering on the set of gradations, such that
${\bf s}\succeq{\bf s}'$ if $s_i\neq0$ whenever $s_i'\neq0$.

We shall sometimes deal with the larger algebra $\gg\oplus{\Bbb C}d$,
formed by
adjoining a derivation with $[d,d]=0$\note{The
algebras formed by adjoining different derivations are equivalent
because $d_{\bf s}-d_{{\bf s}'}\in\gg$.}. The important difference
between $\gg\oplus{\Bbb C}d$ and $\gg$, is that the former has an
invariant
symmetric {\it non-degenerate\/} bi-linear form $(\cdot|\cdot)$,
whereas for the latter the analogous inner-product is degenerate.

In the following we shall be interested in the {\it Heisenberg
subalgebras\/} of $\gg$ [\Ref{KACB},\Ref{KP}],
$\ss={\Bbb C}c+\sum_{j\in E}{\Bbb C}b_j$,
where $E=I+{\Bbb Z}N$, where $I$ is a set of $r$ integers $\geq0$ and
$<N$, for an integer $N$, the algebra being
$$
[b_j,b_k]=j\delta_{j,-k}c.
\efr
For each Heisenberg subalgebra there is an associated gradation ${\bf
s}'$ and derivation $d_{{\bf s}'}$ such that
$$
[d_{{\bf s}'},b_j]=jb_j.
\efr
The integer $N$ is given by $N=\sum_{i=0}^rk_is_i'$, where $k_i$ are the
{\it Kac labels} ($\sum_{i=0}^{r} a_{ij} k_j =0$).

Integrable
highest weight modules of $\gg$ are defined in terms of a highest
weight vector $v_{\bf s}$, labelled by a gradation $\bf s$ of $\gg$.
The highest weight vector is annihilated by $\gg_{>0}({\bf s})$, so
$$
e_i^+\cdot v_\s=0\qquad\forall i,
\efr
and is an eigenvector of $\gg_0({\bf s})$ with eigenvalues
$$\eqalign{
h_i\cdot v_{\bf s}&=s_iv_{\bf s}\cr
e^-_i\cdot v_{\bf s}&=0\qquad\forall i\ {\rm with}\ s_i=0\cr
d_{\bf s}\cdot v_{\bf s}&=0.\cr}
\efr

The eigenvalue of the centre $c$ on the representation $L({\bf
s})$ is known as the level $k$:
$$
c\cdot v_{\bf s}=\sum_{i=0}^rk_i^\vee h_i\cdot v_{\bf
s}=\left(\sum_{i=0}^rk_i^\vee s_i\right)v_{\bf s},
\efr
hence $k=\sum_{i=0}^rk_i^\vee s_i$. In particular, the integrable
highest weight
representations with $k=1$ are known as the {\it basic\/} representations.

We shall use the notation $v_i=v_\s$, where $s_j=\delta_{ij}$, for the
highest weight vectors of the fundamental representations.

Below we prove a lemma.

\sjump\sjump
\noindent
{\bf Lemma A.1} {\it Given two gradations $\s$ and $\s'$, such
that $\s'\succeq\s$, the highest weight vector $v_\s$ is an
eigenvector of the subalgebra $\gg_0(\s')$.}

\sjump
\noindent
{\it Proof\/}. The proof follows from the fact that if $\s'\succeq\s$ then
$\gg_0(\s')\subset\gg_0(\s)$. But $v_\s$ is an eigenvector of $\gg_0(\s)$ and
hence also of $\gg_0(\s')$. $\square$.

\sjump\sjump
\noindent
{\bf Corollary}. {\it If $\s'\succeq\s$ then $v_\s$ is the unique
vector in $L(\s)$ with lowest $\s'$-grade.}

\sjump
\noindent
{\it Proof\/}. Suppose the converse was true, so there exists
another vector $\phi\neq v_\s$ with the same $\s'$-grade as $v_\s$. This would
require $\phi=a\cdot v_\s$, with $a\in{\cal U}(\gg_0(\s'))$ (the
universal enveloping algebra of $\gg_0(\s')$). But by Lemma A.1,
$v_\s$ is an eigenstate of $\gg_0(\s')$, hence $\phi\propto v_\s$
contrary to the hypothesis.
$\square$

\sjump\sjump
Associated to each Kac-Moody algebra $\gg$, there is a group $G$
formed by exponentiating the action of $\gg$ (see [\Ref{PK}] for details).
We denote by $U_{\pm}(\s)$ and $H(\s)$ the subgroups formed by
exponentiating the
subalgebras $\gg_{>0}(\s)$, $\gg_{<0}(\s)$ and $\gg_0(\s)$,
respectively. The group acts projectively on the representations.

\references

\beginref
\Rref{IM}{H-J. Imbens, {\sl Drinfel'd-Sokolov hierarchies and
$\tau$-functions\/}, in: {\sl Infinite dimensional Lie algebras and
groups\/}, World Scientific Adv. Ser. in Math. Phys. {\bf7} (1989) 352}
\Rref{GEN1}{M.F. de Groot, T.J. Hollowood and J.L. Miramontes, {\sl
Generalized Drinfel'd-Sokolov Hierarchies\/}, Commun. Math. Phys.
{\bf145} (1992) 57}
\Rref{GEN2}{N.J. Burroughes, M.F. de Groot, T.J. Hollowood and J.L.
Miramontes, {\sl Generalized Drinfel'd-Sokolov hierarchies II: the
Hamiltonian structures\/}, Princeton Preprint PUPT-1263, IASSNS-HEP-91/42}
\Rref{GEN3}{N.J. Burroughes, M.F. de Groot, T.J. Hollowood and J.L.
Miramontes, {\sl Generalized $W$-algebras and integrable
hierarchies\/}, Phys. Lett {\bf B277} (1992) 89}
\Rref{KP}{V.G. Kac and D.H. Peterson, {\sl 112 constructions of the
basic representation of the loop group of $E_8$\/}, in: Symp. on
Anomalies, geometry and topology, eds. W.A. Bardeen and A.R. White.
World Scientific, Singapore, 1985}
\Rref{KW}{V.G. Kac and M. Wakimoto, {\sl Exceptional hierarchies of
soliton equations\/}, Proceedings of Symposia in Pure Mathematics.
Vol 49 (1989) 191}
\Rref{HIR}{R. Hirota, {\sl Direct methods in soliton theory\/}, in:
{\sl Soliton\/} page 157: eds. R.K. Bullough and P.S. Caudrey (1980)}
\Rref{JAP}{M. Jimbo and T. Miwa, {\sl Solitons and infinite Lie
algebras\/}, Publ. RIMS, Kyoto Univ. {\bf19} (1983) 943\newline
E. Date, M. Jimbo, M. Kashiwara and T. Miwa, {\sl Transformation
groups for soliton equations. Euclidean Lie algebras and reduction of
the KP hierarchy\/}, Publ. RIMS, Kyoto Univ. {\bf18} (1982) 1077}
\Rref{KACB}{V.G. Kac, {\sl Infinite dimensional Lie algebras\/},
$3^{\rm rd}$ edition. Cambridge University Press 1990}
\Rref{PK}{D.H. Peterson and V.G. Kac, {\sl Infinite flag varieties
and conjugacy theorems\/}, Proc. Nat. Acad. Sci. U.S.A. {\bf80} (1983)
1778}
\Rref{DS}{V.G. Drinfel'd and V.V. Sokolov, {\sl Lie algebras and
equations of the Korteweg-de Vries type\/}, J. Sov. Math. {\bf30}
(1985) 1975}
\Rref{WI}{G. Wilson, {\sl Infinite-dimensional Lie groups and algebraic
geometry in soliton theory\/}, Phil. Trans. R. Soc. Lond. {\bf A315}
(1985) 393; {\sl Habillage et fonctions $\tau$\/}, C. r. hebd. S\'eanc.
Acad. Paris {\bf 299} (I) (1984) 587}
\Rref{SW}{G. Segal and G. Wilson, {\sl Loop groups and equations of
Korteweg-de Vries type\/}, Inst. Hautes \'Etudes Sci. Publ. Math.
{\bf63} (1985) 1}
\Rref{VIR}{R. Dijkgraaf, E. Verlinde and H. Verlinde, {\sl Loop
equations and Virasoro constraints in non-perturbative 2D quantum gravity\/},
Nucl. Phys. {\bf B348} (1991) 435\newline
M. Fukuma, H. Kawai and R. Nakayama, {\sl Continuum Schwinger-Dyson
equations and universal structures in two-dimensional quantum
gravity\/}, Int. J. Mod. Phys. {\bf A6} (1991) 507\newline
V.G. Kac and A. Schwarz, {\sl Geometrical interpretation of the
partition function of 2D gravity\/}, Phys. Lett. {\bf B257} (1991)
329}
\Rref{WCON}{M. Fukuma, H. Kawai and R. Nakayama, {\sl Infinite
dimensional Grassmannian structure of two-dimensional gravity\/},
Commun. Math. Phys. {\bf143} (1992) 403\newline
J. Goeree, {\sl $W$-constraints in 2D quantum gravity\/}, Nucl. Phys.
{\bf B358} (1991) 737}
\Rref{US}{T.J. Hollowood, J.L. Miramontes and J. S\'anchez Guill\'en,
{\sl in preparation\/}}
\Rref{VERTEX}{I.B. Frenkel and V.G. Kac, {\sl Basic representations of
affine algebras and dual resonance models\/}, Invent. Math. {\bf62}
(1980) 23\newline
J. Lepowsky and R.L. Wilson, {\sl Construction of the
affine Lie algebra\/}, Commun. Math. Phys. {\bf62} (1978) 43\newline
V.G. Kac, D.A. Kazhdan, J. Lepowsky and R.L. Wilson, {\sl Realization
of the basic representation of the Euclidean Lie algebras\/}, Adv. in
Math. {\bf42} (1981) 83\newline
J. Lepowsky, {\sl Calculus of twisted vertex operators\/}, Proc. Natl.
Acad. Sci. USA {\bf82} (1985) 8295}
\endref
\ciao
